\documentclass[12pt,preprint]{aastex}
\usepackage{psfig,natbib}
\newcommand{\ra}{\mbox{\,$\rightarrow$}}
\newcommand{\Msun}{\mbox{\,$\rm{M_{\odot}}$}} 
\newcommand{\Rsun}{\mbox{\,$\rm{R_{\odot}}$}} 
\newcommand{\Lsun}{\mbox{\,$\rm{L_{\odot}}$}} 

\newcommand{\Xsun}{\mbox{\,$\rm{X_{\odot}}$}}
\newcommand{\Tstar}{\mbox{\,$T_{*}$}}
\newcommand{\Teff}{\mbox{\,$T_{eff}$}}
\newcommand{\Rstar}{\mbox{\,$R_{*}$}} 
\newcommand{\Reff}{\mbox{\,$R_{2/3}$}} 
\newcommand{\Rtrans}{\mbox{\,$R_{t}$}} 

\newcommand{\Minit}{\mbox{\,$M_{init}$}} 
\newcommand{\Mdot}{\mbox{\,$\dot{M}$}}

\newcommand{\Mdotc}{\mbox{\,$\dot{M}_{C}$}}
\newcommand{\Mdots}{\mbox{\,$\dot{M}_{S}$}}

\newcommand{\vturb}{\mbox{\,$v_{turb}$}}

\newcommand{\vinf}{\mbox{\,v$_{\infty}$}}
\newcommand{\logg}{\mbox{\,$\log{g}$}}
\newcommand{\logN}{\mbox{\,$\log{N}$}}
\newcommand{\E}[1]{\mbox{\,$\rm x 10^{#1}$}}
\newcommand{\XH}{\mbox{\,$X_{H}$}}
\newcommand{\XHe}{\mbox{\,$X_{He}$}}
\newcommand{\XC}{\mbox{\,$X_{C}$}}
\newcommand{\XN}{\mbox{\,$X_{N}$}}
\newcommand{\XO}{\mbox{\,$X_{O}$}}
\newcommand{\XSi}{\mbox{\,$X_{Si}$}}
\newcommand{\XS}{\mbox{\,$X_{S}$}}

\newcommand{\XFe}{\mbox{\,$X_{Fe}$}}

\newcommand{\Htwo}{\mbox{\rm{H}$_2$}}
\newcommand{\HI}{\mbox{\rm{\ion{H}{1}}}}

\newcommand{\HeII}{\ion{He}{2}}

\newcommand{\CIII}{\ion{C}{3}}
\newcommand{\CIV}{\ion{C}{4}}

\newcommand{\NIV}{\ion{N}{4}}
\newcommand{\NV}{\ion{N}{5}}

\newcommand{\OIII}{\ion{O}{3}}
\newcommand{\OIV}{\ion{O}{4}}
\newcommand{\OV}{\ion{O}{5}}
\newcommand{\OVI}{\ion{O}{6}}

\newcommand{\SiIV}{\ion{Si}{4}}

\newcommand{\PV}{\ion{P}{5}}

\newcommand{\SV}{\ion{S}{5}}
\newcommand{\SVI}{\ion{S}{6}}
\newcommand{\SVII}{\ion{S}{7}}

\newcommand{\FeIV}{\ion{Fe}{4}}
\newcommand{\FeV}{\ion{Fe}{5}}
\newcommand{\FeVI}{\ion{Fe}{6}}
\newcommand{\FeVII}{\ion{Fe}{7}}
\newcommand{\FeVIII}{\ion{Fe}{8}}
\newcommand{\FeIX}{\ion{Fe}{9}}
\newcommand{\FeX}{\ion{Fe}{10}}

\newcommand{\eg}{\emph{e.g.}}
\newcommand{\ie}{\emph{i.e.}}
\newcommand{\doublet}{$\lambda\lambda$}
\newcommand{\singlet}{$\lambda$}

\newcommand{\EBMV}{\mbox{\,$E_{\rm{B-V}}$}}

\newcommand{\Rv}{\mbox{\,$R_v$}}

\newcommand{\Hbeta}{H$\beta$}

\newcommand{\Lya}{Ly$\alpha$}
\newcommand{\Lyb}{Ly$\beta$}

\newcommand{\Telec}{\mbox{\,$T_{e}$}}

\newcommand{\kms}{\mbox{\,$\rm{km\:s^{-1}}$}}
\newcommand{\gunit}{\mbox{\,$\rm{cm\:s^{-2}}$}}
\newcommand{\Msunyr}{\mbox{\,$\rm{M_{\odot}\:yr^{-1}}$}}

\received{2003 December 23}
\accepted{2004 March 9}
\begin{document}

\title{FAR-UV SPECTROSCOPIC ANALYSES OF FOUR
  CENTRAL STARS OF PLANETARY NEBULAE\footnote{Based on observations made with
the NASA-CNES-CSA Far Ultraviolet Spectroscopic Explorer and data from
  the MAST archive. FUSE is
operated for NASA by the Johns Hopkins University under NASA contract
NAS5-32985.}}

\author{J.E. Herald, L. Bianchi}
\vspace{1mm}
\affil{Department of Physics and Astronomy, The Johns Hopkins University}
\authoraddr{3400 N. Charles St., Baltimore, MD 21218-2411}

\begin{abstract}
We analyze the Far-UV/UV spectra of four central stars of planetary
nebulae with strong wind features --- NGC~2371, Abell~78, IC~4776 and
NGC~1535,  and derive their photospheric and wind parameters by
modeling high-resolution FUSE (Far-Ultraviolet Spectroscopic Explorer)
data in the Far-UV and HST-STIS and IUE data in the UV with spherical
non-LTE line-blanketed model atmospheres.  Abell~78 is a hydrogen-deficient
transitional [WR]-PG~1159 object, and we find NGC~2371 to be in the
same stage, both migrating from the constant-luminosity phase to the
white dwarf cooling sequence with $\Teff \simeq 120$~kK, $\Mdot \simeq
5\E{-8}$~\Msunyr.  NGC~1535 is a ``hydrogen-rich'' O(H) CSPN, and the
exact nature of IC~4776 is ambiguous, although it appears to be helium
burning.  Both objects lie on
the constant-luminosity branch of post-AGB evolution and have $\Teff
\simeq 65$~kK, $\Mdot \simeq 1\E{-8}$.  Thus, both the H-rich and
H-deficient channels of PN evolution are represented in our sample.
We also investigate the effects of including higher ionization stages
of iron (up to \FeX) in the model atmosphere calculations of these hot
objects (usually neglected in previous analyses), and find iron to be
a useful diagnostic of the stellar parameters in some cases.  The
Far-UV spectra of all four objects show evidence of hot ($T \sim
300$~K) molecular hydrogen in their circumstellar environments.

\end{abstract}

\keywords{stars: atmospheres --- stars: Wolf-Rayet --- stars:
  individual (NGC~2371, Abell~78, NGC~1535, IC~4776) --- planetary
  nebulae: general --- UV: spectroscopy}

\section{INTRODUCTION}\label{sec:intro}

After leaving the asymptotic giant branch (AGB), low and
intermediate-mass stars become central stars of planetary nebulae
(CSPN).  The spectral types of the central stars are numerous and
diverse (\eg, O, Of, sdO, [WR], [WR]-PG~1159, as well as white
dwarf-types), but all of these can be generally divided into two
groups based on surface abundances: hydrogen-rich (which show obvious
hydrogen lines in their spectra) and hydrogen-deficient (which don't).
H-deficient stars are in a post-helium flash,
helium-burning phase, while the H-rich stars may be either hydrogen or
helium burning.  About 10-20~\% of CSPN are H-deficient
(\citealp{demarco:02,koesterke:98b} and references therein).  The two
groups are thought to represent different ``channels'' of CSPN
evolution, terminating with either H-rich DA type or He-rich DO type
white dwarfs (\eg, \citealp{napiwotzki:99}).  During the CSPN phase
the star sheds mass in the form of a stellar wind as it evolves toward
the white dwarf cooling sequence.  It is generally believed that the
winds of these objects are driven by radiation pressure, as with
population-I O-stars and WR stars.  If the driving of the wind is
mainly dictated by the stellar luminosity, the wind parameters can, in
principle, be used to determine the distances to CSPN
\citep{kudritzki:99,kudritzki:00,tinkler:02}.

The spectra of the majority of the H-deficient class have strong wind
signatures, show no photospheric lines and are very similar to those
of WC stars --- the carbon-rich Wolf-Rayets which are evolved massive
stars.  The low-mass central stars are termed ``[WC]'' stars to denote
their association with planetary nebulae.  ``PG~1159''- type stars are
extremely hot, H-deficient white dwarf-type objects, thought to
represent an entry point onto the white dwarf cooling sequence.  They
show mainly absorption lines in the optical spectra but possess a few wind
lines in the UV.  Because PG~1159 objects have similar abundances to
[WC] stars and about half of the PG~1159 stars are CSPN
\citep{werner:01}, they are thought to be descendants of the [WC]s.
Linking these two classes is a rare group of objects named
``[WC]-PG~1159'' stars, whose spectra show both emission lines and
some absorption lines.  Only three definite cases of such stars are
known so far: Abell~78, Abell 30 and Longmore 4 \citep{koesterke:01}.
\citet{parthasarathy:98} have posited that all ``weak emission line
stars'' ([WELS]) are actually [WC]-PG~1159, based on optical spectra,
but \citet{werner:01} cautioned against this statement until better
spectra of the sample is available.  The H-rich CSPN sample of
\citet{mendez:88} are similar in that their spectra also feature absorption
lines as well as UV wind emission features.

[WC], [WC]-PG~1159, and PG~1159 stars all reside in the same area of
the H-R diagram (HRD).  In terms of evolution, this region corresponds to
the bend where the star leaves the constant-luminosity post-AGB phase
and is transitioning to the white dwarf cooling sequence. [WC] and
PG~1159 stars lie close to each other in the H-R diagram, yet have
very different spectra (in the optical, a [WC] spectrum is littered
with strong emission lines, while that of a PG~1159 presents an
absorption line spectrum).  This suggests a dramatic drop in mass-loss
rates over a short time of evolution \citep{koesterke:98b}.  The
commonly accepted evolutionary sequence for H-deficient CSPN is:
[WC]\ra [WC]-PG~1159\ra PG~1159\ra DO WD \citep{hamann:96}.  This
sequence spans the time where the stellar winds are in the process of
``turning-off''.  Therefore, understanding the objects in this part
of the H-R diagram is not only a prerequisite for understanding
the evolution of central stars, but will also give insight into the
driving of stellar winds in general.

In the classification of [WR]-type objects, wind parameters are more
fundamental discriminators than stellar temperature
\citep{crowther:99a, acker:03}.  As the central star's winds die down,
the wind features in the FUV/UV are the last to fade
\citep{koesterke:98a}.  Therefore, to estimate accurate parameters of
the extended atmospheres this phase is best studied in the FUV/UV.  In
this paper, we present a FUV/UV-based spectral analysis of four CSPN
which show strong wind signatures - NGC~2371, Abell~78, IC~4776, and
NGC~1535.  We model the FUV/UV spectra of these stars using stellar
atmospheres codes to determine the wind parameters such as \Teff,
\Mdot, and \vinf\ and discuss their evolutionary implications.
Abell~78 has been classified as a [WO]-PG~1159 transition object
\citep{mendez:91,crowther:98}, and we argue that NGC~2371 also belongs
to this class.  NGC~1535 is an O(H) star belonging to the H-rich
sample of \citet{mendez:88}. The FUV spectra of IC~4776 is similar to
that of NGC~1535, but we find evidence that this object may be a
He-burner.  Comparing with evolutionary calculations, our derived
parameters place the latter two objects along the constant-luminosity
path of the HRD, while the former two lie on the bend marking the
transition from that phase to the white dwarf cooling sequence.

This paper is arranged as follows.  The observations and data
reduction are described in \S~\ref{sec:obs}.  A comparison of the
spectra of the objects is presented in \S~\ref{sec:description}.  Our
models and parameter determinations are described in
\S~\ref{sec:modeling}.  The implications of our results are discussed
in \S~\ref{sec:discussion} and our conclusions in
\S~\ref{sec:conclusions}.

\section{OBSERVATIONS AND REDUCTION}\label{sec:obs}

The coordinates, radial velocities ($v_r$) and nebular diameters
($\theta$) of our four sample objects are listed in
Table~\ref{tab:coords} (\citealp{tylenda:03} has measured two angular
dimensions for three of the sample PN, corresponding to the
semi-major and minor diameters, which we list as
``$\theta_B \times \theta_A$'').  The data sets utilized in this paper
are summarized in Table~\ref{tab:obs}.  NGC~2371 and IC~4776 were
observed as part of FUSE cycle 1 program P133 (Bianchi).  All other FUV/UV
data were retrieved from the MAST archive, including FUSE and STIS
archive data of NGC~1535, a Berkeley Extreme and Far-UV Spectrometer
(BEFS) archive spectrum of Abell~78, and IUE data of most objects.
The reduction of the FUSE data is described in \S~\ref{sec:fuse}.

\subsection{FUV Data}\label{sec:fuse}

FUSE covers the wavelength range of 905--1187~\AA\ at a spectral
resolution of $\lesssim$ 30,000. It is described by \citet{moos:00} and
its on-orbit performance is discussed by \citet{sahnow:00}. FUSE collects
light concurrently in four different channels (LiF1, LiF2, SiC1, and
SiC2).  Each channel is recorded by two detectors, each divided into
two segments (A \& B) covering different subsets of the above range
with some overlap.

The FUSE spectra were taken using the LWRS
(30\arcsec$\times$30\arcsec) aperture.  These data, taken in
``time-tag'' mode, have been calibrated using the most recent FUSE
data reduction pipeline, efficiency curves and wavelength solutions
(CALFUSE v2.2).  We combined the data from different segments, weighting
by detector sensitivity, and rebinned them to a uniform dispersion of
0.05~\AA\ (which is probably close to the actual resolution since the
data were taken in the early part of the mission).  Bad areas of the
detectors, and those regimes affected by an instrumental artifact
known as ``the worm'' (FUSE Data Handbook v1.1), were excluded.  The
flux calibration accuracy of FUSE is $\lesssim 10$~\%
\citep{sahnow:00}.

The BEFS spectra covers a wavelength range of 912--1218~\AA\ with a
resolution of $\sim 0.33$~\AA.  The calibration and performance are
described by \citet{hurwitz:98}.

\subsection{UV Data}

For most stars, we make use of IUE data, using high-resolution
spectrum if available (a high-resolution spectra for NGC~2371 exists,
but is of poor quality, so we present the low-resolution data
instead).  For some stars, there was some disagreement ($\lesssim
30$~\%) between the FUV (both FUSE and BEFS) and IUE continuum levels
in the regions of overlap (flux calibrations for these instruments are
typically accurate to $\sim10$~\%).  Because there are no nearby
objects that could contribute extra flux in the FUSE aperture, we
suspect that the background has been over-subtracted in the IUE spectra
since nebular continuum emission is expected (in the case of NGC~1535,
the STIS spectra agree perfectly well with the FUSE fluxes, but the
IUE fluxes are a factor of $\sim1.2$ lower).  Therefore we we have
scaled the UV flux levels to those of the FUV (scaling
factors are listed in Table~\ref{tab:obs}).

\section{DESCRIPTION OF THE SPECTRA}\label{sec:description}

\begin{figure}
\begin{center}
\epsscale{.90}
\rotatebox{180}{
\plotone{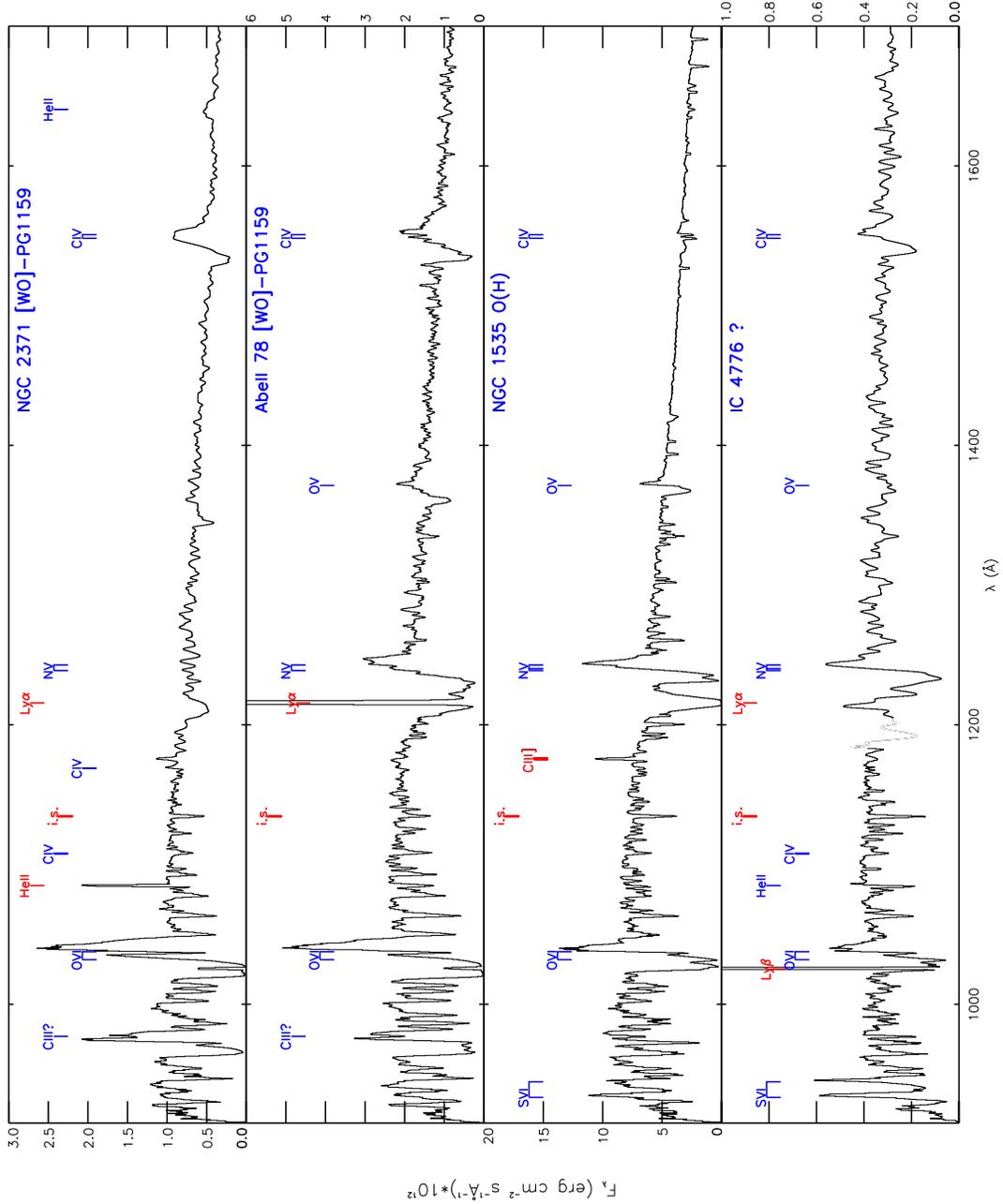}}
\caption{FUV/UV spectra of our sample.  Data are from different
  instruments (with different resolutions - see Table~\ref{tab:obs}).
  The spectrum of IC~4776 appears flatter than the others because of
  the contribution from the nebular continuum (see
  \S~\ref{sec:nebcont}).  Prominent stellar features are marked with
  dark/blue labels, and interstellar/nebular features by light/red
  features. Regions of bad data are blanked out.
  }\label{fig:uv}
\end{center}
\end{figure}

Fig.~\ref{fig:uv} displays the combined FUV/UV spectra of our sample
(up to 1700~\AA, longwards of this, the spectra are mostly
featureless).  Much of the fine structure between 900---1200~\AA\ is
caused by sight-line molecular hydrogen (\Htwo), and is discussed in
\S~\ref{sec:htwo}.  As far as stellar features go, the spectra present
a sparse landscape.  The \OVI\ \doublet 1032-38 feature is most
prominent in all cases, and the spectra are arranged from top to bottom 
in order of decreasing strength of this feature.
IC~4776 and NGC~1535 both display  \SVI\
\doublet 933-44 in the FUV, however, their UV spectra are different.  In
the FUV, Abell~78 and NGC~2371 have 3 similar lines but \NV\ \doublet
1238-43 and \OV\ \singlet 1371 are strong P-Cygni in Abell~78 while
absent in NGC~2371, neither showing \SVI\ but both displaying a
P-Cygni feature around 977~\AA.  This feature is commonly assumed to
be \CIII, however the formation of a \CIII\ line in the highly-ionized
winds of these objects is hard to explain, and this will be discussed
later (\S~\ref{sec:discussion}).

\CIV\ \doublet 1548-51 appears in all spectra, although it is
relatively weak in NGC~1535.  \NV\ \doublet 1238-43 is obviously
present in all spectra save that of NGC~2371.  \OV\ \singlet 1371 is
absent in NGC~2371, present as a P-Cygni profile in Abell~78 and
NGC~1535, and appears perhaps only in absorption in IC~4776.  \HeII\
\singlet 1640 appears to be present in the low-resolution IUE spectrum
of NGC~2371, but inspection of the one available (noisy)
high-resolution IUE spectra for this object (SWP53006) shows the
\HeII\ \singlet 1640 to be at least partly nebular in origin.  The
absence of transitions from ions of lower ionization potential (\eg,
\NIV, \OIV, or \SiIV) indicate the high ionization state of the winds,
and high effective temperatures of the central stars.

\section{MODELING}\label{sec:modeling}

The modeling of these spectra consists of three parts: modeling the
central star spectrum, the nebular continuum emission longwards of
1200~\AA, and the sight-line Hydrogen (atomic and molecular) absorption
shortwards.  The reddening must also be adjusted concurrently to match
the FUV/UV slope.  In all four objects, the reddening is low, and not
a significant source of uncertainty.
Iterations between each are required to achieve a
consistent solution.  However, we discuss each in turn for clarity.

\subsection{The Central Stars}\label{sec:cs}

The optical spectrum of these objects shows mainly absorption lines,
and some have been modeled using plane-parallel codes (\eg, Abell~78
by \citealp{werner:03}, NGC~1535 by \citealp{mendez:88}).  In the
FUV/UV, the most prominent features are wind lines.  The temperatures
and wind parameters of some of these objects have been previously
determined from plane-parallel analyses in conjunction with wind-line
analyses of one or two wind features (such as \CIV\ \doublet 1548-51
or \OVI\ \doublet 1332-38 --- \citealp{koesterke:98b}).  Here we
derive wind parameters solely from the FUV/UV spectra.

Intense radiation fields, a (relatively) low wind density, and an
extended atmosphere invalidate the assumptions of thermodynamic
equilibrium and a plane-parallel geometry for the winds of our sample CSPN.
To model these winds, we use the non-LTE (NLTE) line-blanketed code
CMFGEN of \citet{hillier:98,hillier:99b}.  CMFGEN solves the radiative
transfer equation in an extended, spherically-symmetric expanding
atmosphere.  Originally developed to model the Wolf-Rayet winds, it
has been adapted for objects with weaker winds such as O-stars and
CSPN as described in \citet{hillier:03}.  The detailed workings of the
code are explained in the references therein.  To summarize, the code
solves for the NLTE populations in the co-moving frame of reference.
The fundamental photospheric/wind parameters include \Teff, \Rstar,
\Mdot, the elemental abundances and the velocity law (including
\vinf).  The \emph{stellar radius} (\Rstar) is taken to be the inner
boundary of the model atmosphere (corresponding to a Rosseland optical
depth of $\sim20$).  The temperature at different depths is determined
by the \emph{stellar temperature} \Tstar, related to the luminosity
and radius by $L = 4\pi\Rstar^2\sigma\Tstar^4$, whereas the
\emph{effective temperature} (\Teff) is similarly defined but at a
radius corresponding to a Rosseland optical depth of 2/3 (\Reff).  The
luminosity is conserved at all depths, so $L =
4\pi\Reff^2\sigma\Teff^4$. 

The terminal velocity (\vinf) can be estimated from the blue edge of
the P-Cygni absorption features, preferably from features that extend
further out in the wind (we used \CIV\ \doublet 1548-51 primarily).
We assume what is essentially a standard velocity law $v(r) =
\vinf(1-r_0/r)^\beta$ where $r_0$ is roughly equal to \Rstar. 
Once a velocity law is specified, the density structure
of the wind is then parameterized by the mass-loss rate through the
equation of continuity: $\Mdot=4\pi\Rstar^2 \rho(r)v(r)$.

It has been found that models with the same \emph{transformed radius}
\Rtrans\ [$\propto \Rstar(\vinf/\Mdot)^{2/3}$] \citep{schmutz:89} and
\vinf\ have the same ionization structure, temperature stratification
(aside from a scaling by \Rstar) and spectra (aside from a scaling of
the absolute flux by $\Rstar^2$ --- \citealp{schmutz:89,hamann:93}).
Thus, once the velocity law and abundances are set, one parameter may
be fixed (say \Rstar) and parameter space can then be explored by
varying only the other two parameters (\eg, \Mdot\ and \Teff).
\Rtrans\ can be thought of as an optical depth parameter, as the
optical depth of the wind scales as $\propto \Rtrans^{-2}$, for
opacities which are proportional to the square of the density.
Scaling the model to the observed flux yields \Rstar/$D$ --- a
distance $D$ must be adopted to determine \Rstar.


For the model ions, CMFGEN utilizes the concept of ``superlevels'',
whereby levels of similar energies are grouped together and treated as
a single level in the rate equations \citep{hillier:98}.
Ions and the number of levels and superlevels included in the model
calculations, as well references to the atomic data, are given in the
Appendix (\S~\ref{sec:atomic}).

\subsubsection{Clumping}\label{sec:clumping}

Radiation driven winds have been shown to be inherently unstable
\citep{owocki:88,owocki:94}, which should lead to the formation of
clumps in the expanding atmosphere.  The degree of clumpiness is
parametrized by $f$, the \emph{filling factor}.  One actually can only
derive \Mdots=(\Mdotc/$\sqrt{f}$) from the models, where \Mdots\ and
\Mdotc\ are the smooth and clumped mass-loss rates.  For the denser
winds of population-I Wolf-Rayet stars, the clumping factor can be
constrained by the strength of the electron scattered optical line
wings, and clumping factors of $f \sim 0.1$ are typical (a reduction
of \Mdot\ to roughly a third of its smooth value).  For O-stars, the
lower mass loss rates make the electron scattering effects small and
difficult to constrain \citep{hillier:03}.  The winds of our sample
stars are even weaker, and we do not attempt to determine the clumping
for our sample.  Unless otherwise noted, we have adopted $f=1$ and use
\Mdot\ to refer to the smooth mass loss rate throughout this paper.
This can be seen as an upper limit.  One expects the clumping in the
more tenuous CSPN wind to be less severe than in a WR wind, so a
conservative estimate of the lower limit of \Mdot\ would be a third of
this value.

\subsubsection{Gravity}\label{sec:gravity}

Because of the FUV/UV range contains mostly wind lines, there are no
suitable absorption lines to be used as gravity diagnostics.  In
CMFGEN, gravity enters through the scale height $h$ ($\propto
g^{-1}$), which connects the spherically extended hydrostatic outer
layers to the wind.  The relation between $h$ and $g$, defined in
\citet{hillier:03}, involves the mean ionic mass and mean number of
electrons per ion, the local electron temperature and the ratio of
radiation pressure to the gravity.  \citet{mendez:88} determined
$\logg = 4.3\pm0.2$ for NGC~1535 and \citet{werner:03} found
$\logg=5.5$ for Abell~78 through analysis of photospheric optical
lines.  We initially adopted $\logg = 4.3$ for NGC~1535 and
IC~4776, and $\logg = 5.5$ for Abell~78 and NGC~2371.  For such
gravities, the derived wind parameters are not that sensitive to $h$.
Spectral wind features can be fit with the models of the same \Teff\
but with gravities differing by over a magnitude.  We discuss the
issue of gravity more in \S~\ref{sec:discussion}.

\subsubsection{Abundances}\label{sec:abund}

Throughout this work, the nomenclature $X_i$
represents the mass fraction of element $i$, ``\Xsun'' denotes the
solar abundance, and our values for ``solar'' are taken from
\citet{gray:92}.

Because of the lack of photospheric lines, we have made assumptions
about the abundances of the model atmospheres.  We have computed
models with two different abundance patterns: solar abundance for all
elements (corresponding to H-rich objects) and carbon/oxygen enriched
``[WC]-PG~1159''-type abundances, typical of H-deficient CSPN.  With
respect to the latter, it has been shown that abundances through the
[WC] subclasses are about the same \citep{crowther:02,crowther:98},
with the spectroscopic differences mainly tied to \Mdot\ and \Teff\
\citep{crowther:02}.  Calculations by \citet{herwig:01} show that these
abundances may result when the star experiences a late He-shell flash.
The abundances of PG~1159 objects are similar
\citep{gorny:00}.  Typical measured values are (by mass): \XHe =
0.33--0.80, \XC = 0.15--0.50, \XO = 0.06--0.17 \citep{werner:01}.  For
these reasons, we have assumed a similar abundance pattern for these
elements (\XHe/\XC/\XO = 0.54/0.36/0.08), while adopting solar
abundances for others (\eg, S, Si, \& Fe).  The nitrogen abundance in
such objects typically ranges from undetectable to $\sim2$\% (by mass,
see \citealp{werner:01} and references therein).  We have adopted a
nitrogen abundance of 1\% for most of our models.  The lone \NV\
\doublet 1238-43 feature is abundance sensitive, and we do not rely on
it strongly as a diagnostic as it is also sensitive to \Teff\ and
\Mdot.  We use these abundances in our models of Abell~78 (a
[WR]-PG~1159 star) and of NGC~2371, which has a similar FUV spectrum
(Fig.~\ref{fig:uv}).

For the H-rich O(H) star NGC~1535, we have adopted solar abundances
for all elements.  For IC~4776, which displays a similar FUV/UV
spectrum, we have computed models with both solar abundances and the above
H-deficient ``[WC]/PG~1159'' abundances.  We are able to fit its
spectrum adequately with models of both abundances (we will discuss
this more in \S~\ref{sec:results}).

Table~\ref{tab:abund} shows our final model elemental abundances.  For
most, our assumed abundances produced
adequate fits.  Some required adjusting, as described in
\S~\ref{sec:results} in a case-by-case basis.

\subsubsection{Diagnostics}\label{sec:diag}

Optimally, one determines \Teff\ by the ionization balance of He or
the CNO elements.  It is desirable to use diagnostics from many
elements to ensure consistency.  However, in the FUV/UV spectra of
these objects only oxygen shows two ionization stages -- \OV\ and
\OVI.  We find the \OVI\ feature to be mainly insensitive to \Mdot\
and moderately sensitive to \Teff, however, the possible influence of
X-rays on this line make it less than desirable temperature diagnostic
(in the O-stars regime, \OVI\ is also sensitive to \Mdot, but above a
sharp threshold, and this line is primarily created by X-rays ---
\citealp{bianchi:02}).  The other elements (C, N, S) only show one
ionization stage.  Because of the lack of multiple ionization stages
from the same element in the spectra of most of our sample, effective
temperatures are mainly derived from the appearance or absence of ions
of different ionization potentials.  For example, in our cooler
objects (IC~4776 and NGC~1535) the absence of low ionization features
such as \SV (73 eV), \PV (65 eV), \OIV (77 eV), and \NIV (77 eV) places
a lower limit on \Teff\ ($\sim$55--60~kK from our models) for a wide
range of mass-loss rates and abundances. \citet{tinkler:02} find that
between 55-70~kK, H-rich CSPN exhibit a large jump in \Mdot\ by a
factor of 10-100, which they suspect is due to a bi-stability jump.
In our models, it is in this temperature range that these previously
mentioned ions jump to the next stage, supporting the idea of
\citet{vink:01} that a bi-stability jump at such high temperatures may
be due to CNO elements.  The presence of \SVI\ in these object's
spectra further constrains the temperature to be below a certain
point.  Above $\Teff \gtrsim 90$~kK, \SVII\ becomes dominant in the
outer wind, and a regime is entered in which our ``hotter'' objects
lie (NGC~2371 and Abell~78).  Around \Teff$\simeq$120~kK, the winds
become too ionized for \OV, a discriminator between our two hot
objects.

Saturated features (like \OVI) are mainly insensitive to the mass-loss
rate.  We typically use the observed \CIV, \OV, and \SVI\ lines to
constrain \Mdot.  Also, \HeII\ \singlet 1640 (not observed in most of
our sample objects) is used to place upper limits on this parameter.
In some cases, we also use the iron spectrum as an additional diagnostic
(\S~\ref{sec:iron}).

\subsubsection{Iron}\label{sec:iron}

The effects of iron (and other line-blanketing elements) are important
to consider in the modeling of hot stars for multiple reasons.  
It is known that in the modeling of hot stars, the neglect of
important line-blanketing elements such as iron can significantly
impact the derived parameters.  Furthermore, iron features may be used
to provide additional diagnostics of stellar parameters (\eg,
\citealp{herald:01}).  Finally, line-blanketing elements such as iron
are high sources of opacity in the wind and are thus important in the
understanding of how it is driven.

Our sample represents an ideal test-bed for studying the effects of
the inclusion of higher ionization stages of iron (\ie, \FeVII--\FeX) in
the model atmospheres, which have typically been neglected in previous
studies of similar objects.  The temperatures of the objects in our
sample span a range where different stages in this sequence dominate.
Since \FeIV--\FeX\ have transitions which occur in the FUV and UV, 
the wavelength coverage of our spectral data is adequate to study
these effects.

\begin{figure}[htbp]
\begin{center}
\epsscale{1}
\rotatebox{0}{
\plotone{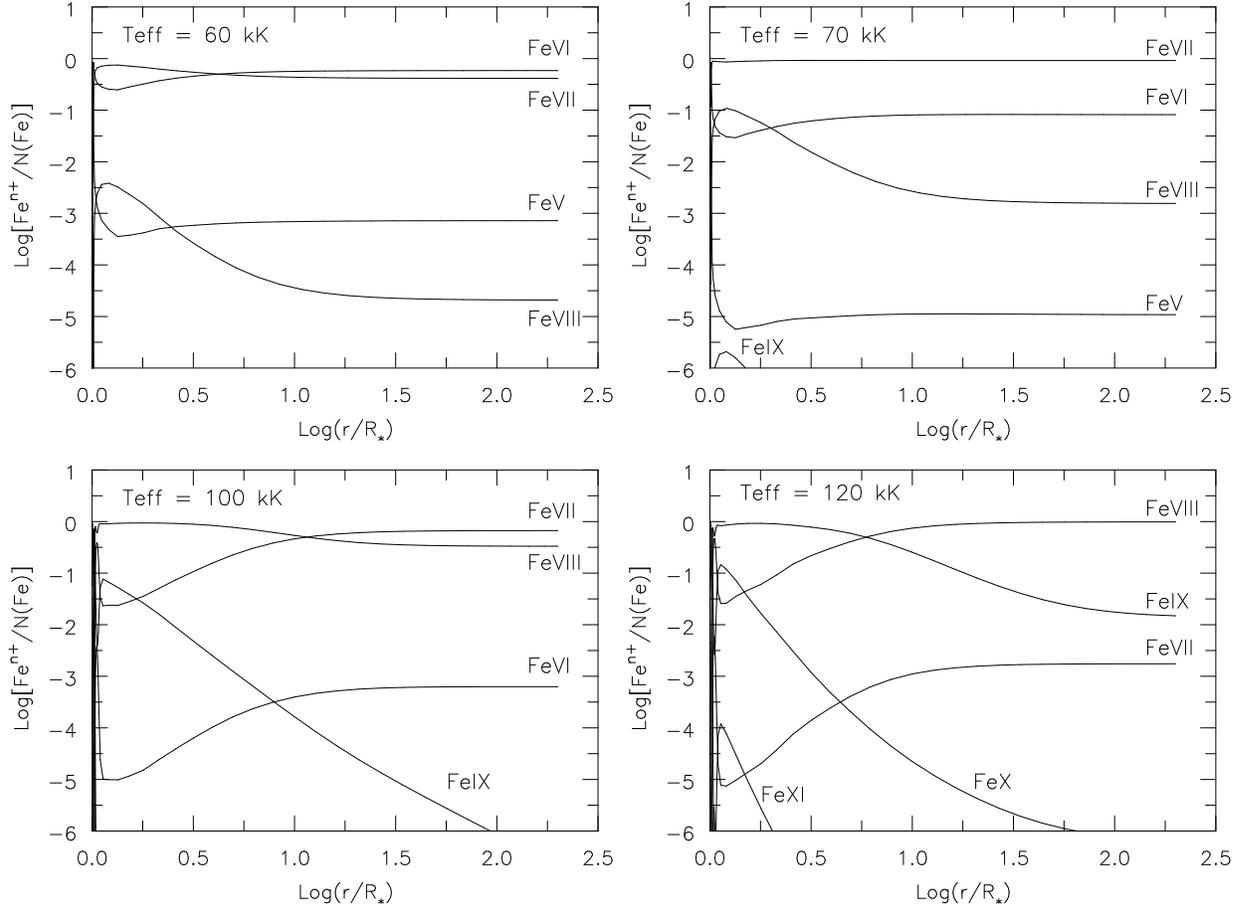}}
\caption{The ionization structure of iron for models spanning the
  temperature ranges of our objects, as a function of distance from
  the stellar surface.  The upper panels are models with
  temperatures of our cooler objects ($\Teff \simeq 60-70$~kK), where
  \FeVI--\FeVII\ dominate throughout the wind.  The lower panels
  models in the temperature range of our hotter objects ($\Teff \simeq
  100-130$~kK), where the higher ionization stages are dominant.
  These figures illustrate the sensitivity of the ionization structure
  of iron to \Teff.}\label{fig:ion_iron}
\end{center}
\end{figure}

\begin{figure}[htbp]
\begin{center}
\epsscale{.28}
\rotatebox{270}{
\plotone{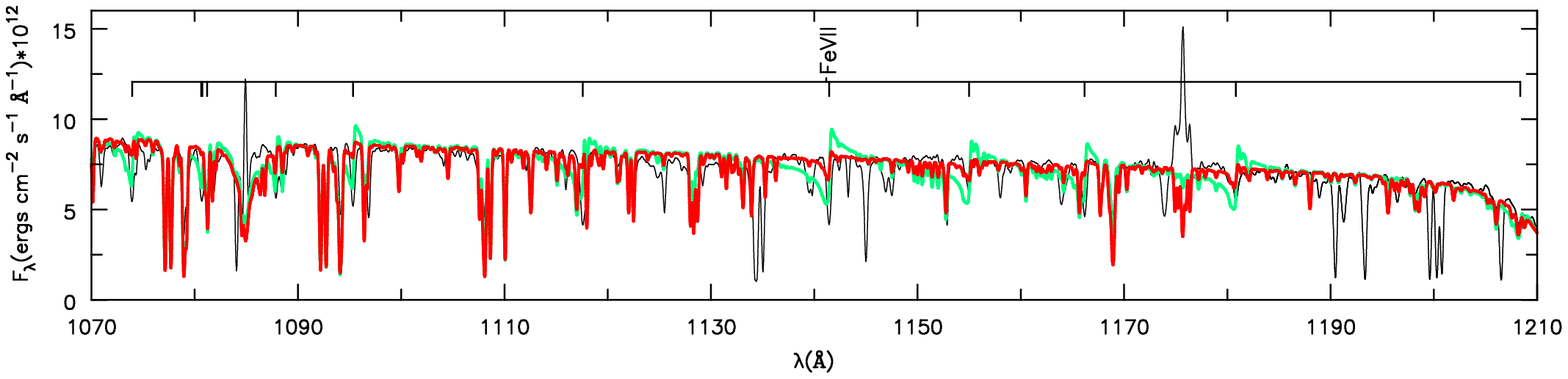}}
\caption{NGC~1535: Iron spectrum as a
diagnostic. Observations of NGC~1535 are shown (black) along with
a model with $\Teff\simeq75$~kK, $\Mdot \simeq 3\E{-8}$~\Msunyr\
(green/light gray).  This model displays an \FeVII\ spectrum stronger
than the observations, indicating a lower mass-loss rate and/or a
different temperature is required.  Our final model, with $\Teff
\simeq 65$~kK, $\Mdot\simeq1\E{-8}$~\Msunyr\ (red/heavy gray)
provides a better fit.}\label{fig:iron_1535}
\end{center}
\end{figure}

The ionization structures of iron for models in the temperature range
of our sample (with similar mass-loss rates) are shown in
Fig.~\ref{fig:ion_iron}.  The figures illustrate what ionization
stages are dominant in various temperature regimes.  For the cooler
objects, \FeVI--\FeVII\ dominate.  As \Teff\ increases, \FeVIII\
becomes dominant throughout the wind (between $\Teff \gtrsim 100$~kK)
and at the highest temperatures ($\Teff \gtrsim 120$~kK) \FeIX\
becomes important in the inner winds layers.  Useful iron diagnostics
include the array of \FeV\ transitions occurring between
1350--1500~\AA, and the forest of \FeVI--\FeVII\ lines between 1250 and
1500~\AA.  Fig.~\ref{fig:iron_1535}
demonstrates how the iron spectrum can be used as a diagnostic of wind
parameters for NGC~1535.

\subsubsection{Stellar Modeling Results}\label{sec:results}

\begin{figure}[htbp]
\begin{center}
\epsscale{.9}
\rotatebox{180}{
\plotone{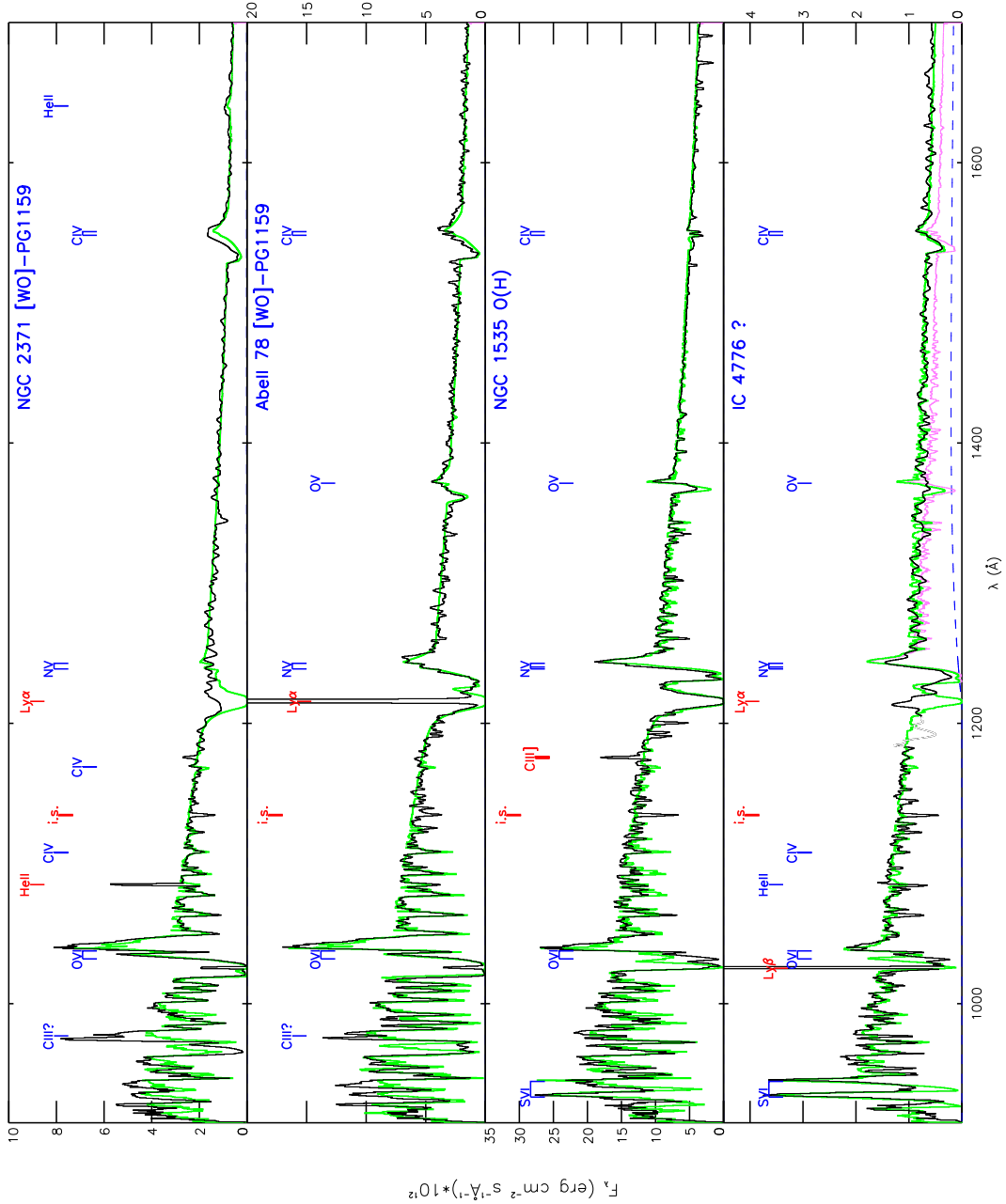}}
\caption{Model fits.  The observations (black) have been unreddened
  using our values for \EBMV\ derived from the UV slope, and our
  models (green/thick gray) are also shown.  In the case of IC~4776,
  the model nebular continuum (blue/dark dashed), and the stellar
  model (pink/light gray) are summed together (green/thick gray).  The
  prominent wind features are marked with black labels, and strong
  nebular and interstellar features are marked with gray/red
  labels. The spectra have been convolved with a 0.75~\AA\ Gaussian
  for clarity.  }\label{fig:mod}
\end{center}
\end{figure}

Our model fits are shown in Fig.~\ref{fig:mod}.
We now discuss our results for the individual objects, along with some
notes.  Our derived stellar parameters and a global discussion is
given in \S~\ref{sec:discussion}.

\noindent \emph{NGC~1535}:
The strongest wind features (\SVI, \OVI, \NV, \OV, \CIV) of this O(H)
star are all well fit by our model using solar abundances.  As
discussed in \S~\ref{sec:iron}, the iron spectrum provides an
additional constraint for this O(H) star, as at higher temperatures
the \FeVII\ features are too strong (reducing \Mdot\ serves to weaken
the non-iron wind features to unacceptable levels).

Our derived parameters ($\Teff = 66\pm 5$~kK, $\log{\Mdot} =
-8.10\pm0.3$~\Msunyr) agree within the uncertainties with those of
\citet{koesterke:04}, who model this object using a different wind
code ($\Teff = 70$~kK, $\log{\Mdot} = -7.8$~\Msunyr).  The temperature
is about the same as that found by \citet{mendez:88} in their
plane-parallel analysis of optical spectra ($\Teff=58\pm5$~kK, $\logg
= 4.3\pm0.2$).  It is cooler than that derived by \citet{tinkler:02}
who reported the following parameters for NGC~1535: $\Teff = 80$~kK,
$\Rstar=0.38$~\Rsun, $\log{L}=3.76$~\Lsun, $\logg=5.05$,
$\vinf=1900$~\kms, and $\log{\Mdot}=-8.7\pm0.7$~\Msunyr\ ($\Rtrans =
430^{+570}_{-280}$).  Their mass-loss rate, which is about a factor of
4 lower than ours, was derived solely from \OV\ \singlet 1371, which
we find is quite sensitive to \Teff\ in this temperature range.

\noindent \emph{IC~4776}: The nebular continuum for this object
strongly contributes to the observed UV spectrum at longer wavelengths
(\S~\ref{sec:nebcont}).  We are able to achieve fits of about the same
quality for this object using either our H-rich or H-deficient
abundances by adjusting \Teff\ within the range 55--70~kK and
$\log{\Mdot}$ within $-7.85\pm0.30$~\Msunyr, with the H-rich models
yielding slightly better fits (it is our H-rich model which we present
in the figures).  \OVI\ 1032-38 is a bit weak at $\Teff \simeq 60$,
but fits well at $\Teff=70$~kK.  The \SVI\ 933-44 absorption is too
strong in our presented model.  It is better fit with the lower limit
of our mass-loss rate, but the other wind features are then a bit
weak.  Our models of either H-rich and H-deficient abundances
overproduce \OV\ \singlet 1371, which is weak at best in the
observations.  Cooler models bring in (unobserved) \OIV\ features and
lower mass-loss rates weaken all other features.  Reducing the oxygen
abundance to 1/100th solar brings the feature in line with
observations, but then the \OVI\ doublet in the FUV range is far too
weak.  \citet{bouret:03} had difficulties fitting this feature in O
dwarf stars, which have mass-loss rates of order
$10^{-9}-10^{-6}$~\Msunyr\ and are cooler ($\Teff \simeq 40-50$~kK).
They found a very clumped wind ($f = 0.01$) alleviated the problem
somewhat.

Our mass-loss rates also agree with the predictions we obtained using
the \citet{vink:01} prescription (\S~\ref{sec:discussion}), which uses
normal (\ie, non-He or C enriched) abundances.  The low mass-loss rate
of these objects and the dearth of diagnostics make the abundances
difficult to constrain, and from our analysis, we cannot tell if
IC~4776 is H-rich or H-deficient.  Comparison with the evolutionary
tracks (also see \S~\ref{sec:discussion}) indicate this objects is
most probably a He-burner.

IC~4776 has been classified as a [WC6] star \citep{feibelman:99} and
as a ``weak emission lined object'' (a WELS --- \citealp{tylenda:93}).
The [WC6] classification for IC~4776 was based on the equivalent width of
\CIII\ \singlet 2297.  This classification is questionable for the
following reasons.  The spectra of a [WC5] or [WC6] object should
contain relatively strong \CIII\ features throughout its spectrum
(\eg, \citealp{crowther:98,hillier:99b}), whereas the FUV/UV spectrum
of IC4776 is relatively featureless.  The FUSE spectrum of this object
shows high ionization features such as \OVI\ and \SVI, and
indicating a state of higher ionization than one would expect in a
[WC6] object.

\noindent \emph{Abell~78}:
The prominent wind lines of this object (\OVI, \NV, \OV, \CIV) are all
well-reproduced by our model, with the exception
of the strong P-Cygni feature at the location of \CIII\ \singlet 977.
\citet{koesterke:98b} also had problems reproducing this feature in
Abell~78, and speculated that neglected iron lines might sufficiently
cool the outer layers of the (otherwise hot) wind to allow for the
formation of \CIII.  However, we have tested models accounting for
ionization stages of iron up to \FeX\ and had no success (the dominant
ionization stages in the model winds for these objects are \FeVIII\
and \FeIX).  As iron is likely depleted in this object (see below), this is
probably not the solution (although cooling from other species is
possible). 

Using a solar value for the iron abundance ([Fe]=7.67), we could not
simultaneously fit the iron spectrum and our other diagnostic lines.
For any appreciable mass-loss rate, relatively strong (unobserved)
iron P-Cygni features are seen in models with temperatures spanning
$\sim$70---130~kK.  For lower temperatures ($\Teff \lesssim 95$~kK),
the set of \FeVII\ lines shown in Fig.~\ref{fig:iron_1535} are seen.
The hotter models display strong \FeVIII\ and/or
\FeIX\ features.  This supports
the findings of \citet{werner:03}, who found Abell~78 to have an iron
abundance of at most $\sim$0.03 solar.  Using this iron abundance, the
model spectrum agrees with the observations.  \citet{miksa:02} have
also found iron deficiency in a large sample of PG~1159 stars, the
supposed descendants of these transitional objects.  Iron deficiencies
in these objects may result when material in the He-intershell is
exposed to $s$-process nucleosynthesis during a thermally pulsating
AGB or post-AGB phase  \citep{lugaro:03,herwig:03}.

\citet{koesterke:01} determined $\Tstar=115$~kK, $\Rtrans=45.2$~\Rsun\
and $\vinf=3750$~\kms\ for this object, with mass fractions for C/O/N
of 40/15/2~\%, respectively.  Our parameters are essentially the same
(36/8/1), except we derive $\vinf=3200\pm50$ from the fitting the
\CIV\ \doublet 1248-51.  \citet{werner:03} derived $\Teff=110$~kK and
$\logg=5.5$, also in line with our determinations.

Abell~78 is the prototype `` [WC]-PG~1159'' star
\citep{hamann:96,koesterke:98b} and was originally placed in a
transition phase between the [WR] and PG~1159 [\ie, O(C)] phases by
\citet{mendez:91}.  \citet{crowther:98} suggested a [WO1]-PG~1159
classification based its high \OVI\ \singlet 3818/\CIV\ \singlet 5808
ratio. This object also falls into the category termed ``\OVI -type
CSPN'', distinguished by \citet{smith:69} from the [WC] class for
having \OVI\ \doublet 3811-34 as one of their most prominent optical
features. Our model parameters (derived solely on the FUV and UV
spectral region) adequately reproduces the optical wind features of
Abell~78.

Abell~78 is also of interest because it belongs to a small group of PN
which have H-poor, dusty ejecta.  These nebulae consist of a H-poor
shell surrounded by a an outer H-rich region, indicating a 2nd ejection
event of H-deficient material during the post-AGB phase
\citep{medina:00}.

\noindent \emph{NGC~2371}: As can be seen in Fig.~\ref{fig:uv}, the
FUV flux of this object is almost identical to that of Abell~78,
except the wind lines are more conspicuous and broader, indicating a
higher terminal velocity and mass-loss rate.  The UV spectrum lacks
\OV, which, in our models, disappears for $\Teff \gtrsim
$115---120~kK.  There is a weak signature of \NV\ 1238-43.  Because of
the low resolution of our data, it is unclear if its origin is nebular
or stellar.  Assuming the latter, this feature weakens enough to
fit the observations in our synthetic spectra for models with $\Teff
\gtrsim 125$~kK (it should be kept in mind we have assumed a mass
fraction of $\XN =0.01$ for nitrogen).  The iron spectrum further
constrains the temperature to be $\Teff \gtrsim 130$~kK, because the
\FeIX\ features in the FUV are too strong for lower \Teff.  However,
if one assumes that this object is iron-deficient as is Abell~78 (see
\S~\ref{sec:results}), this restriction is removed.  As with Abell~78, our
models fail to reproduce the strong wind feature at 977~\AA.

The low resolution IUE spectrum shows \HeII\ \singlet 1640, but the
one high resolution IUE spectrum available shows that at least part of
the \HeII\ emission is nebular.  Our upper limit to the mass-loss rate
assumes it is all stellar, and our lower limit assumes no stellar
\HeII\ line.  The high-resolution spectrum also reveals the \CIV\
emission to be partly nebular.  The absorption trough appears to be
saturated (the high-resolution spectrum is under-exposed, and hence
extremely noisy, but around the stronger lines there is enough flux to
see these details).

\citet{tylenda:93} classified NGC~2371 as a [WC3].  However, we find
the ionization of the wind too high for such a classification.  Our
model parameters for NGC~2371 are close to those of Abell~78
(PG~1159-[WO1]), with the former star being a bit hotter with a denser
wind.  Their synthetic optical spectra are qualitatively similar, and
so we suggest a [WO]-PG~1159 classification for NGC~2371, but
at a higher state of ionization marked by the absence of \OV\ and the
presence of \OVI.  This classification needs to be confirmed with
optical spectra, however.

\citet{kaler:93} noted a peculiarity with the optical \OVI\ emission
features of this star: they consisted of two components, a broad,
blended feature (with \vinf$\simeq 3400$~\kms) and two resolved
narrower features.  It is the only known \OVI\ PNN to display both the
broad and narrow \OVI\ features. They believed these narrow features
were associated with the wind rather than the nebula, because their
long-slit CCD spectrum showed no extension of the narrow \OVI\
features beyond the profile of the CS, confining the \OVI\ formation
zone to have a radius of $\sim$2000 AU. 
It appears the \OVI\ \doublet 1032-38 for NGC~2371 also has
this two component structure, as shown in Fig.~\ref{fig:OVI}.  Close
inspection reveals a narrow emission line superimposed on top of the
broad P-Cygni profile, which could correspond to the blue component of
the \OVI\ doublet (at $\sim$ 20~\kms\ in the rest wavelength of the
star).  The red feature may also be present, but \Htwo\ absorption
prevents a firm conclusion.  These narrow \OVI\ components 
may be evidence of shocked material in the nebula.  \citet{herald:04a}
found similar (seemingly nebular) \OVI\ features in the LMC CSPN SMP
LMC~62.  

In their study of \OVI\ PNN, \citet{stanghellini:95} have found a
correlation between the strength of \OVI\ \doublet 3811-3834 and
stellar luminosity.  According to the models of our two transition
stars, this holds true.  Throughout our sample, the \OVI\ 1032-38
feature grows stronger with increasing effective temperature and luminosity.

\subsection{Molecular and Atomic Hydrogen}\label{sec:htwo}

Absorption due to atomic and molecular Hydrogen (\Htwo) along the
sight-line complicates the spectra of these objects in the FUSE range
where numerous \Htwo\ transitions from the Lyman ($B^1
\Sigma^+_u$--$X^1 \Sigma^+_g$) and Werner ($C^1 \Pi^{\pm}_u$--$X^1
\Sigma^+_g$) sequences lie.  Toward a CSPN, this sight-line Hydrogen
typically consists of interstellar and circumstellar components.
Material comprising the circumstellar \HI\ and \Htwo\ presumably was
ejected from the star earlier in its history (during the AGB phase),
and is thus important from an evolutionary perspective.  If hot (\eg,
$T\gtrsim300$K), a small column density [\eg,
$\log(N)\sim16$~cm$^{-2}$] of \Htwo\ can lead to a complex absorption
spectrum, obscuring the underlying stellar spectrum (see
\citealp{herald:04a}, Fig.~5).  It is therefore necessary to model the
\HI\ and \Htwo\ characteristics to discern the underlying stellar
spectrum.  Our parameter determinations for sight-line \HI\ and \Htwo\
are listed in Table~\ref{tab:htwo}.

\begin{figure}[htbp]
\begin{center}
\epsscale{.28}
\rotatebox{270}{
\plotone{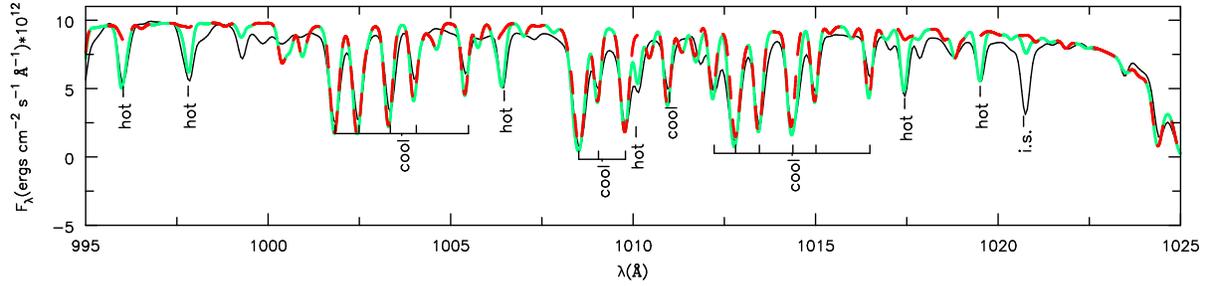}}
\caption{Fitting \Htwo: The FUSE spectrum of NGC~1535 is shown (black)
along with a stellar model with successive hydrogen absorption
components applied (Table~\ref{tab:htwo}).  The dashed red/dark gray line
shows the stellar model with absorption from the ``cool'' ($T =
80$~kK) \HI\ and \Htwo\ components applied, and the green/light gray
line with absorption effects of the ``hot'' \Htwo\ component ($T =
400$) \Htwo\ applied to that.  Features attributable to the hot and
cool components, as well as strong interstellar absorption features,
are labeled.  The cool hydrogen absorption model is not sufficient to
fit all the features, suggesting a hot \Htwo\ gas associated with the
nebula.  }\label{fig:h2fit}
\end{center}
\end{figure}

\Htwo\ absorption effects were applied to the model spectrum in the
following manner.  For a given column density ($N$) and gas
temperature ($T$), the absorption profile of each line is calculated
by multiplying the line core optical depth ($\tau_0$) by the Voigt
profile $H(a,x)$ (normalized to unity) where $x$ is the frequency in
Doppler units and $a$ is the ratio of the line damping constant to the
Doppler width (the ``b'' parameter).  The observed flux is then
$F_{obs} = \exp{[-\tau_{0}H(a,x)]} \times F_{intrinsic}$.  To fit the
\Htwo\ spectrum of a given object, we first assume the presence of an
interstellar component with $T=80$~K (corresponding to the mean
temperature of the ISM --- \citealp{hughes:71}) and \vturb = 10\kms.
The column density of this interstellar component is estimated by
fitting the strongest transitions.  If absorption features due to
higher-energy \Htwo\ transitions are observed, a second, hotter
(circumstellar) component is added (see example in
Fig.~\ref{fig:h2fit}).  The temperature of the circumstellar component
can be determined by the absence/presence of absorption features from
transitions of different $J$ states, and the column density by fitting
these features.  Iteration between fitting the interstellar and
circumstellar components are performed, as both contribute to the
lower-energy features. We note that our terminology of
``circumstellar'' and ``interstellar'' components is a simplification,
and basically indicate a ``cool'' component (assumed interstellar) and
``hot'' component (assumed circumstellar).  However, the column
density derived for the cooler ``interstellar'' component may also
include circumstellar \Htwo.

In the case of NGC~1535, the \Htwo\ has been previously modeled by
\citet{bowers:95} using $\sim3$~\AA\ resolution Hopkins Ultraviolet
Telescope.  They found
the data to be well fit employing either a one-component model of
$T=300$~K or a two component model with $T=144$ and $500$~K.  Their
upper limit for the circumstellar \Htwo\ column density was
$\log{N(\Htwo)}=18.4$~$\rm{cm}^{-2}$.  This is consistent with our
findings, in which we have fit the \Htwo\
absorption (see \S~\ref{sec:htwo}) using two components, a cool
component [$T=80$~K, $\log{N(\Htwo)}=18.7$~$\rm{cm}^{-2}$] and a hot
component [$T=400$~K, $\log{N(\Htwo)}=16.7$~$\rm{cm}^{-2}$], thanks to
the higher resolution of the FUSE spectra.

\begin{figure}[htbp]
\begin{center}
\epsscale{.5}
\rotatebox{270}{
\plotone{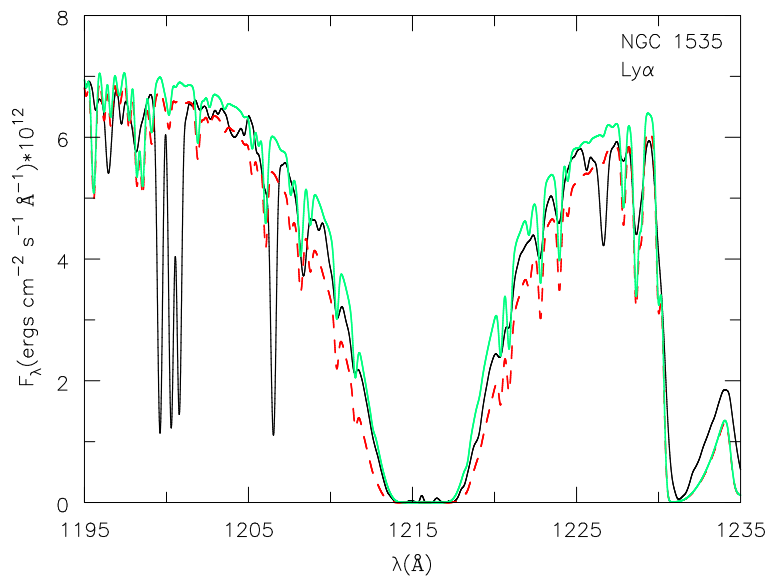}}
\caption{Fitting \Lya\ of NGC~1535: The \Lya\ profile in the STIS data
(black) and our stellar model with the absorption effects of
different amounts of \HI\ applied ($T=80$~kK): column densities of
$\logN=20.8$, (green/light gray) and 21.0~cm$^{-2}$ (dashed red/dark gray).
The models bracket the observations, so the sight-line column density
can be well constrained to $\logN=20.8\pm0.1$~cm$^{-2}$.
}\label{fig:Lya}
\end{center}
\end{figure}

\HI\ column densities were determined in a similar fashion by fitting
the profiles of the \Lya\ and/or \Lyb\ features (our primary
diagnostic feature is indicated in Table~\ref{tab:htwo}) .  For Abell~78 and
NGC~1535, the available high-resolution data of the \Lya\ line enable
the sight-line \HI\ to be constrained tightly, as illustrated in
Fig.~\ref{fig:Lya}.  For IC~4776 and NGC~2371, the low-resolution IUE
spectra are inadequate for this purpose.  For IC~4776, we use the blue
side of the \Lyb\ profile, but for NGC~2371, the P-Cygni absorption
from the \OVI\ doublet extends far enough blueward that \Lyb\ is
obscured.  We therefore simply assumed an \HI\ column density of 21.0
for this object for fitting purposes.

\subsection{Nebular Continuum}\label{sec:nebcont}

Nebular characteristics, taken from the literature, are shown in
Table~\ref{tab:neb}.  Also presented are 
nebular radii ($r_{neb}$) and dynamic (kinematic) ages
$\tau_{dyn} = r_{neb}/v_{exp}$, calculated using the angular sizes
from Table~\ref{tab:coords} and the distances from
Table~\ref{tab:mod_param_dist}.  Since some nebulae have angular
diameter measurements in two dimensions, we used the average, listed
as $\theta_{adopt}$.

To determine if the nebula significantly contributes to the UV flux,
nebular continuum emission models were computed accounting for
two-photon, H and He recombination, and free-free emission processes.
The continuum model parameters are the electron density ($n_{e}$), the
electron temperature ($T_e$), the Helium to Hydrogen ratio (He/H) and
the doubly to singly ionized Helium ratio (He$^{2+}$/He$^+$).  The
computed emissivity coefficient of the nebular gas was scaled as an
initial approximation to the total flux at the Earth by deriving the
emitting volume absolute \Hbeta\ flux F(\Hbeta), dereddened using
$c_{H\beta} = 1.475 \EBMV$.  If the nebula was not entirely contained
within the aperture used for the observations, the continuum was scaled
by an appropriate geometrical factor ($4A/\pi\theta^{2}$ or
$4A/\pi\theta_A \theta_B$ where $A$ is the area of the aperture in
square arcseconds).  Because the value of F(\Hbeta) is very uncertain
the continuum model was then re-scaled to match the observed flux.

\begin{figure}[htbp]
\begin{center}
\epsscale{.25}
\rotatebox{270}{
\plotone{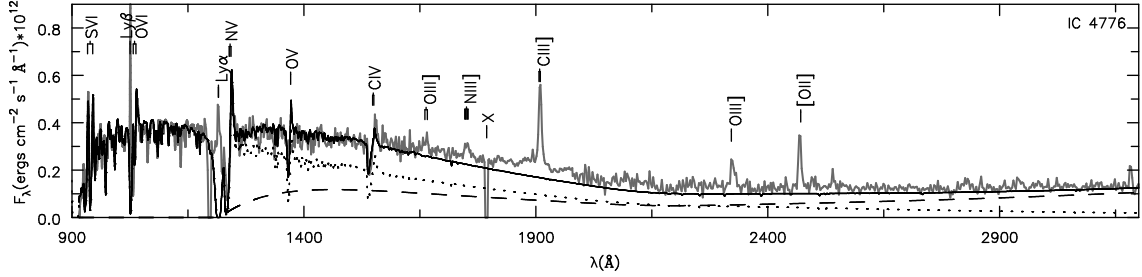}}
\caption{IC~4776: The observations
  are shown (gray) along with our stellar (dotted) and nebular
  continuum models (dashed).  The sum of both models is also shown
  (solid), demonstrating the flux longwards of \Lya\ (1215~\AA) of this
  object is significantly affected by the nebular continuum.  We have
  applied our molecular hydrogen absorption model (\S~\ref{sec:htwo}) to 
  the stellar model, and our atomic hydrogen absorption model to both
  the stellar and nebular models.  In the FUV, the observations and
  models have been convolved with a 2~\AA\ Gaussian for
  clarity.
}\label{fig:ic4776_neb}
\end{center}
\end{figure}

\begin{figure}[htbp]
\epsscale{0.5}
\rotatebox{270}{
\plotone{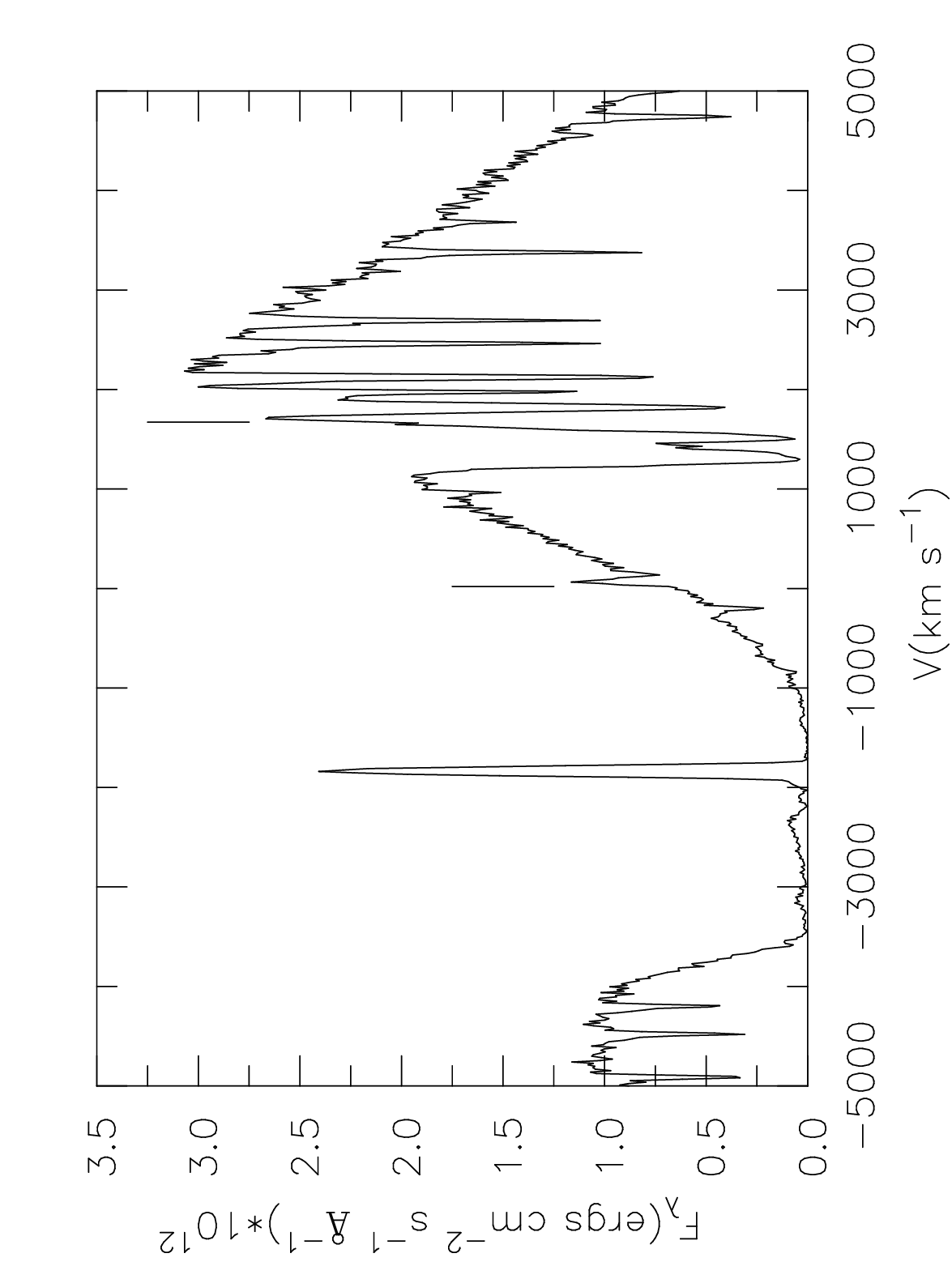}}
\caption{\OVI\ \doublet 1031.93,1037.62 for NGC~2371, in velocity
  space around 1031.93\AA.  The spectra has been blueshifted to
  account for the measured radial velocity of the star of
  $v_r=+20.6$~\kms\ (Table~\ref{tab:coords}).  Marked is a narrow
  emission line at $\sim20$~\kms, probably corresponding to the blue
  component of the doublet.  We have also marked the corresponding
  position of the red component, assuming it is velocity shifted by
  the same amount.  The red component seems to be present, but the
  absorption from \Htwo\ prevents positive identification.
  \citet{kaler:93} noticed similar narrow emission lines on top of the
  broad P-Cygni profiles in optical \OVI\ features.  }\label{fig:OVI}
\end{figure}

For NGC~2371, Abell~78, and NGC~1535, the estimated nebular continuum
fluxes are $\sim1$\%, $\lesssim 0.1$\%, and $\lesssim 0.01$\%,
respectively, of the observed flux between 1400--1500~\AA.  These
fluxes do not significantly affect the modeling of the stellar
spectra.  However, the contribution of the nebular continuum emission
to the UV spectra of IC~4776 is significant, and is responsible for
the shape of the observed spectrum at the longer UV wavelengths.  This
can be seen in Fig.~\ref{fig:ic4776_neb}, which shows the UV spectrum
of this object along with our stellar and nebular continuum models.

\subsection{Reddening}\label{sec:red}

In Table~\ref{tab:red_param}, we list reddenings determined
from the FUV/UV spectral slope, from our measured \HI\ column
densities, and from literature.  We discuss each in turn.

For even the coolest objects of our sample, wavelengths above 912~\AA\
lie in the Rayleigh-Jeans tail of the spectral energy distribution.
This makes the slope of the FUV/UV continuum insensitive to \Teff\ and
mainly dependent on the reddening toward the object.  We have thus
used the slope of the FUV/UV spectra to constrain \EBMV\ toward three
of our objects, for which the nebular continuum contribution is
negligible.  In doing so, we have used the reddening law of
\citet{cardelli:89} assuming $\Rv=3.1$, the standard Galactic value.
Although their work was originally valid only for wavelengths longer
than 1250~\AA, ongoing work has shown that the law may be safely
extended through the FUSE range (G. Clayton 2001, private
communication).  In the case of IC~4776, the nebular continuum
contamination in the UV (\S~\ref{sec:nebcont}) hinders this
method.  However the FUV spectrum is primarily stellar, and the
determination of \EBMV\ was based on this range.  The sum of the
stellar and nebular components, and the derived \EBMV, fit well with
the overall wavelength range (900-3000~\AA).

Our values determined from the UV slope agree well with those
determined from literature values of the logarithmic extinction (at
H$\beta$) using the relation $c_{H\beta} = 1.475 \EBMV$.
In all cases, the reddening is very small, and does not affect the
results of the stellar modeling. 

Finally, we list the reddenings implied by our measured column
densities of \HI\ (\S~\ref{sec:htwo}) using the relationship $\left<
N(\HI)/\EBMV \right> = 4.8\E{21}$~cm$^{-2}$~mag$^{-1}$
\citep{bohlin:78}, which represents typical conditions in the ISM.
This results in significantly higher reddenings than the other two
methods, indicating that the derived column density includes a significant
amount of \emph{circumstellar} \HI, which apparently has a smaller
dust-to-\HI\ ratio than that of the ISM.  This is similar to our
findings for other CSPN \citep{herald:02,herald:04a}.

\section{RESULTS and DISCUSSION}\label{sec:discussion}

As discussed in \S~\ref{sec:cs}, values of \logg\ were taken or
estimated from previous works based on optical spectra (photospheric
lines).  From our spectral fitting process we determined the
distance-independent parameters \Teff, \Rtrans, and \vinf.  Once a
distance is adopted, \Rstar\ was derived by scaling the model flux to the
observations, and then $L$ and \Mdot\ were determined.  Distances to our
objects found in the literature derived by various methods show large
spreads (\eg, for NGC~1535, 0.8---3.1~kpc --- \citealp{sabbadin:84b}).
Most are determined from nebular relations, such as that of the
nebular radius to ionized mass.  \citet{ciardullo:99} have shown that,
in comparison with distance determinations made from more reliable
methods such as spectroscopic parallax, statistical methods typically
are off by a factor of 2 or more.  With this in mind, our adopted
distances from literature are shown in Table~\ref{tab:mod_param_dist},
along with the resulting parameters.

The results of comparing the effective temperatures and luminosities
of our sample to the evolutionary tracks of \citet{vassiliadis:94} are
shown in Table~\ref{tab:mod_param_tracks}.  They include the core mass
($M_c$, essentially the current mass), the initial mass of the
progenitor (\Minit), the evolutionary age ($\tau_{evol}$) and the
gravity derived using $M_c$ and the radii
(Table~\ref{tab:mod_param_dist}).  NGC~1535, the H-rich object, was
compared with the H-burning tracks, and lies along the
constant-luminosity phase.  We used the He-burning tracks for NGC~2371
and Abell~78, which both fall on the bend between the
constant-luminosity phase and the WD cooling sequence, as appropriate
for their transitional nature. At the temperature of IC~4776, the
H-burning tracks extend down to luminosities of $\log{L} \gtrsim
3.5$~\Lsun\ (corresponding to the $M_{init} = 1.0$~\Msun\ track).  The
upper limit of our luminosity of IC~4776 ($\log{L} =
3.2^{+0.27}_{-0.15}$~\Lsun) does not overlap with the H-burning
tracks.  The parameters of IC~4776 do, however, fall on the He-burning
tracks, which we have used in this case.  Which tracks were used
are listed in the Table.

In most cases, masses calculated using our derived radii and initially
adopted gravities ($\logg = 5.5$ for Abell~78 and NGC~2371,
$\logg=4.3$ for NGC~1535 and IC~4776 --- \S~\ref{sec:gravity})
are unreasonably low (\ie, $\lesssim 0.5$~\Msun).
Gravities computed using the track masses are higher than our
adopted values in all cases.  The gravities of NGC~2371 and Abell~78
($\logg=6.3$ and $5.7$, respectively) are good agreement with PG~1159
stars of similar temperatures from the 16-object sample of
\citet{miksa:02}.  Likewise, the gravities of NGC~1535 and IC~4776
($\logg=4.9$ and $5.1$) are reasonable for CSPN of these temperatures
(\eg, \citealp{mendez:88}).  Models with these higher gravities
resulted in only slight changes to the stellar parameters, and within
the errors quotes in Table~\ref{tab:mod_param_dist}.  Therefore we
conclude these higher gravities are more appropriate for our sample
than those originally adopted from the literature.

For most of our objects, the dynamic ages (Table~\ref{tab:neb}) are 2-4 times
lower than the evolutionary ages.  This is not unexpected, as the
dynamic age is a lower limit to the post-AGB age, because the nebular
expansion increases during the early post-AGB phase and then level off
as the nucleus fades (see \citealp{sabbadin:84}, \citealp{bianchi:92}
and references therein).  The exception is Abell~78, which has a
larger dynamic age than its evolutionary age ($\tau_{dyn} = 15$~kyr
vs. $\tau_{evol} = 10$).

\citet{vink:01} presented wind models that yield mass-loss rates for
given stellar parameters, taking into account opacity shifts in the
wind due to different ionization structures at different temperatures.
They assume ``normal'' abundances - \ie, not He or CNO enriched.
Although their prescriptions were derived from studies of the winds of
(massive) O and B stars, we applied their prescription using the
derived parameters (\Teff, $L$, \vinf, and \logg) of our
\emph{low-mass} objects (Table~\ref{tab:mod_param_dist}), using solar
metallicity.  The results are listed in
Table~\ref{tab:mod_param_dist}.  The predicted mass-loss rate of
NGC~1535 (which we fit with normal abundances) is the same as our
measured value.  For NGC~2371 and Abell~78 (the H-deficient objects), the
predicted mass-loss rates are many magnitudes lower than our values.
This is probably due to the chemically enriched winds of these objects
having much higher opacities.  For IC~4776, a predicted mass-loss rate
is a magnitude lower than our measured value results.  This may be an
indication that the wind of IC~4776 has some chemical enrichment,
another indication it may be an He-burner.

The mass-loss rates of the H-deficient objects are about a magnitude
larger than the H-rich object, and have higher terminal velocities.
The more powerful winds of the hotter objects are a consequence of the
hydrogen deficiency of the atmosphere, which increases the opacity and
thus the efficiency in converting radiative momentum flux into wind
momentum flux.  A measure of the efficiency of the wind at converting
radiative momentum flux to wind momentum flux is the ``performance
number'' $\eta = \vinf \Mdot c/L$ \citep{springmann:94}.  The
performance number basically measures how many times, on average, each
photon scatters in the wind.  Performance numbers above unity (the
``single scattering limit'') are typical of stars with chemically
enriched winds such as WR objects.  The performance numbers for our
sample are also listed in Table ~\ref{tab:mod_param_dist}, and show
the H-rich object having the lowest $\eta$, while the two H-deficient
objects are above the single-scattering limit.  IC~4776 has a
performance number of unity, again indicating its wind may be
chemically enriched.

\section{CONCLUSIONS}\label{sec:conclusions}

We have performed a FUV/UV spectral analyses on four CSPN which display
combined absorption and wind line spectra, a sign that the winds of
these objects are in the process of switching off.  Two objects are
H-deficient (NGC~2371 and Abell~78), one is H-rich (NGC~1535), and for
one our analysis is inconclusive (IC~4776).  Thus both channels of the
CSPN phase are represented. We derived the stellar wind parameters of
the sample PN by modeling their FUV and UV spectra, where strong wind
signatures are seen.  NGC~1535 and Abell~78 have been previously
analyzed in the optical using plane-parallel models(\citealp{werner:03}, 
\citealp{mendez:88}), and the latter also with wind models
\citep{koesterke:98b}.  Our parameters are in good agreement with the
previous results.

The most striking feature in all the FUV spectra is the \OVI\ doublet.
In massive O-type stars, this ion originates from X-rays due to shocks
in the radiatively driven winds \citep{bianchi:02}.  We are able to fit
this feature in our sample CSPN without the inclusion of X-rays due to
their higher effective temperatures.

Two objects have $\Teff \simeq 65$~kK, with
$\Mdot\sim10^{-8}$~\Msunyr.  NGC~1535 has been classified as a
Hydrogen-rich O(H) star based on its optical spectrum, and we are able
to match its FUV/UV spectrum using solar abundances.  From the
spectrum of IC~4776, we cannot constrain the abundances, but
comparison with evolutionary tracks suggest this object is a
He-burner.  Our models for these objects show a jump in the ionization
structure of the CNO elements (and S and P) between 55-70~kK.  It is
in this temperature range that \citet{tinkler:02} find that H-rich
CSPN exhibit a large jump in \Mdot\ by an factor of 10-100, which they
suspect is due to such a bi-stability jump.  This supports the idea of
\citet{vink:01} that a bi-stability jump at such high temperatures may
be due to CNO elements.

The two H-deficient objects Abell~78 and NGC~1271 are significantly
hotter ($\Teff>110$~kK), with $\Mdot\sim 10^{-7}$~\Msunyr.  Our
derived parameters for the [W01]-PG~1159 star Abell~78 generally agree
with those of past analyses.  We derive similar parameters for
NGC~2371, and suggest that it is a [WR]-PG~1159 also, but its wind is
more ionized and shows no \OV\ features (a [WO0]-PG~1159?).  We find
evidence of iron deficiency in both of these objects,
supporting the findings of \citet{werner:03} in Abell~78 and
of \citet{miksa:02} in PG~1159 stars.  They lie on the bend in the
theoretical evolutionary tracks between the constant luminosity phase
and the WD cooling sequence, having post-AGB ages of 10-15~kyr.  For
[WC] stars, it seems that as the star evolves away from the post-AGB
phase through the [WC] sequence, the temperature and terminal velocity
of the wind increase as the wind density decreases \citep{acker:03}.

Except for Abell~78, the post-AGB ages predicted by the evolutionary
models are typically 2-4 times lower than their kinematic ages.
However, kinematic ages are lower limits to the post-AGB age (as the
expansion is slower in the initial phase) thus the actual discrepancy
may be smaller.

Our FUV/UV analysis has provided wind parameters for H-rich and
H-deficient CSPN at the stage in post-AGB evolution where the winds
are fading.  However, because a precise determination of the mass is
not possible, the objects do not necessarily represent the same
evolutionary sequence.  This work has also provided information on the
interstellar and circumstellar environment from our measurements of
the column densities and temperatures of \HI\ and \Htwo\ along the
sight-lines.  Our determinations of the \HI\ column density and \EBMV\
imply that the relationship between these two quantities in the
circumstellar (PN) environment of CSPN differs from that of the ISM,
having lower dust-to-gas ratios (probably due to the destruction of
dust by the radiation field).  The high resolution of the FUSE data
allow us to detect hot \Htwo\ associated with the nebulae.  For all
four objects in our sample, a single component of \Htwo\ gas at
typical ISM temperatures (\ie, $T \sim 80$~kK) was not adequate to fit
the absorption spectrum.  A second, hotter ($T\sim300$~kK) component
was necessary, which we assume to be associated with the circumstellar
environment.  With the advent of many FUSE observations of CSPN, it is
becoming apparent that hot circumstellar \Htwo\ is not uncommon CSPN
at quite different evolutionary stages: from old PN with white dwarf
nuclei \citep{herald:02} to young, compact PN \citep{herald:04a}.  The
objects in this paper lie either along the constant luminosity section
of the post-AGB evolutionary tracks, or on the transition bend to the
WD cooling sequence, and thus represent intermediate stages to the
previously mentioned cases.  Given the intense UV radiation fields
emitted by the CSPN, it is likely that the nebular \Htwo\ exists in clumps,
shielded by neutral and ionized hydrogen, as appears to be the case in
the Helix nebula \citep{speck:02}.

\acknowledgements

We thank John Hillier for his help with the CMFGEN code, and
Stephan McCandliss for making his \Htwo\ molecular data available.  We
are indebted to the members of the Opacity Project and Iron Project
and to Bob Kurucz for their continuing efforts to compute accurate
atomic data, without which, this project would not have been feasible.  We
are grateful to the referee, Klaus Werner, for his many constructive
comments and suggestions.  The SIMBAD database was used for literature
searches.  This work has been funded by NASA grants NAG 5-9219
(NRA-99-01-LTSA-029), NAG5-10364 and NAG5-10364.  The BEFS, STIS and
IUE data were obtained from the Multimission
Archive (MAST) at the Space Telescope Science Institute (STScI). STScI is
operated by the Association of Universities for Research in Astronomy,
Inc., under NASA contract NAS5-26555.

\appendix

\section{APPENDIX: MODEL ATOMS}\label{sec:atomic}

Ions and the number of levels and superlevels included in the model
calculations are listed in Table~\ref{tab:ion_tab}.  The atomic data
come from a variety of sources, with the Opacity Project
\citep{seaton:87,opacity:95,opacity:97}, the Iron Project
\citep{pradhan:96,hummer:93}, \citet{kurucz:95}\footnote{See
http://cfa-www.harvard.edu/amdata/ampdata/amdata.shtml} and the Atomic
Spectra Database at NIST Physical Laboratory being the principal
sources.  Much of the Kurucz data were obtained directly from CfA
\citep{kurucz:88,kurucz:02}.  Individual sources of atomic data
include the following: \citet{zhang:97}, \citet{bautista:97},
\citet{becker:95b}, \citet{butler:93}, \citet{fuhr:88},
\citet{luo:89a}, \citet{luo:89b}, Mendoza (1983, 1995, private
communication), \citet{mendoza:95},
\citet{nussbaumer:83,nussbaumer:84}, \citet{peach:88}, Storey (1988,
private communication), \citet{tully:90}, and
\citet{wiese:66,wiese:69}.  Unpublished data taken from the Opacity
Project include: \FeVI\ data (Butler, K.), \FeVIII\ data (Saraph and
Storey) and \FeIX\ and \FeX\ data (C. Mendoza).




\clearpage

\clearpage

\clearpage

\clearpage

\clearpage

\clearpage

\clearpage

\clearpage

\clearpage


\begin{table}[htbp]
\caption{The Program Objects}\label{tab:coords}
\begin{tabular}{cccccc}
\hline
Name & PN G & R.A. & Dec & $v_r^a$ & PN Diameter \\
     &     & (J2000) & (J2000) & (\kms) & (\arcsec)\\
\hline
NGC2371 & 189.1+19.8 & 07 25 34.71 & +29 29 26.20 & $+20.6\pm2.7$ & 48.9$\times$30.6$^b$\\
Abell~78 & 081.2-14.9 & 21 35 29.8  & +31 41 40  & $+17\pm10.0$ &107$^a$ \\
NGC~1535 & 206.4-40.5 & 04 14 15.9 & -12 44 21 & $-3.2\pm1.4$ & 33.3$\times$32.1$^b$\\
IC 4776 & 002.0-13.4 & 18 45 50.57 & -33 20 32.94 & $+18.9\pm0.7$ &8.5$\times$4.0$^b$\\
\hline
\end{tabular}
\begin{minipage}{\textwidth}
\begin{trivlist}
\item a: \citet{cahn:71}
\item b: \citet{tylenda:03}, two dimensions given.
\end{trivlist}
\end{minipage}

\end{table}

\begin{table}[htbp]
\caption{Utilized Spectra}\label{tab:obs}
\begin{tabular}{cccccccl}
\hline
Star & Instrument & Data- & Date & Resolution & Aperture & Range & Scale\\
     &      & set    &      &        & (\arcsec)&  (\AA) & Factor \\
\hline
NGC~2371 & FUSE & P1330301 & 02/26/00 & $\sim 0.05$~\AA & 30$\times$30 & 915--1180 &  \\
         & IUE & SWP04883 & 04/07/79  & 5--6~\AA & 10$\times$20 & 1180--1975 & $\times1.3$\\
         & IUE & LWR04210 & 04/07/79  & 5--6~\AA & 10$\times$20 & 1975--3345 & $\times1.3$\\
Abell~78 & BEFS & BEFS2190 & 12/01/96 & $\sim 0.33$~\AA& 20 & 915-1222 & \\
         &  IUE & SWP19906 & 05/05/83 & $\sim 0.2$~\AA & 10$\times$20 & 1122--1975 & $\times1.2$\\
NGC~1535 & FUSE & P1150808 & 10/05/01 & $\sim 0.05$~\AA & 30$\times$30 & 915--1180 &  \\
         & STIS+E140M & O64D04010& 03/01/01 & $\sim 0.05$~\AA &0.2$\times$0.2 & 1180--1700 &\\
IC~4776  & FUSE & P1330501 & 05/21/00 & $\sim 0.05$~\AA & 30$\times$30 & 915--1180 &  \\
         & IUE  & SWP16504 & 03/11/82 & 5--6~\AA & 10$\times$20 & 1180--1975 &$\times1.25$\\
         & IUE & LWR12764  & 03/11/82 & 5--6~\AA & 10$\times$20 & 1975--3345 & $\times1.25$\\
\hline
\end{tabular}
\end{table}

\begin{table}[htbp]
\caption{Model abundances (\Xsun\ = solar abundance)}\label{tab:abund}
\begin{tabular}{ccccccccc}
\hline
Parameter  & \XH & \XHe & \XC & \XN & \XO & \XSi & \XS & \XFe \\
\hline
NGC~2371   & - & 0.54 & 0.37 & $\lesssim$0.01? & 0.08 & \Xsun & \Xsun & \Xsun?\\
Abell~78   & - & 0.54 & 0.37 & 0.01 & 0.08 & \Xsun & \Xsun & $\lesssim$0.03\Xsun\\
NGC~1535   & \Xsun & \Xsun & \Xsun & \Xsun & \Xsun & \Xsun & \Xsun & \Xsun \\
IC~4776    & \Xsun & \Xsun & \Xsun & \Xsun & \Xsun & \Xsun & \Xsun & \Xsun \\
\hline
\end{tabular}
\begin{minipage}{\textwidth}
\begin{trivlist}
\item ?: Value uncertain, see text.
\end{trivlist}
\end{minipage}
\end{table}

\begin{table}[htbp]
\caption{\Htwo\ and \HI\ parameters}\label{tab:htwo}
\begin{tabular}{lllrl}
\hline
Star & Component & $\log{N}$ & $T$ & Note \\
\hline
NGC~2371 & \HI (IS+circ)& 21.0 & 80 & Not fit \\
        & \Htwo\ (IS) & 18.0$^{+0.7}_{-0.3}$ & 80 & \\
        & \Htwo\ (circ) & 17.0$_{-0.7}^{+0.3}$ & 300 &\\
Abell~78 & \HI (IS+circ)& 21.1$_{-0.2}^{+0.2}$ & 80 & From \Lya\\
        & \Htwo\ (IS) & 19.7$_{-0.7}^{+0.3}$ & 80 & \\
        & \Htwo\ (circ) & 16.4$_{+0.3}^{-0.7}$ & 300 & \\
NGC~1535 & \HI (IS+circ)& 20.8$\pm$0.1 & 80 &  From \Lya \\
        & \Htwo\ (IS) & 18.7$_{+0.3}^{-0.7}$ & 80 & \\
        & \Htwo\ (circ) & 16.4$_{-0.4}^{+0.3}$ & 400 & \\
IC~4776 & \HI (IS+circ)& 21.4$_{-0.4}^{+0.3}$ & 80 & From \Lyb \\
        & \Htwo\ (IS) & 15.7$_{+0.3}^{-0.7}$ & 80 &  \\
        & \Htwo\ (circ) & 15.7$_{+0.3}^{-0.7}$ & 300 &  \\
\hline
\end{tabular}
\end{table}
\begin{deluxetable}{lccccccccc}
\rotate
\tabletypesize{\footnotesize}
\tablecolumns{9}
\tablewidth{0pc}
\tablecaption{Nebular Parameters and Estimated Kinematic Ages\label{tab:neb}}
\tablehead{
\colhead{Name} & 
\colhead{$n_{e}$} & 
\colhead{\Telec} & 
\colhead{$\log(F_{H\beta}^{obs})$} &
\colhead{He/H} & 
\colhead{He$^{2+}$/He$^+$} &
\colhead{$v_{exp}$(\OIII)} &
\colhead{$\theta_{adopt}$} &
\colhead{$r_{neb}$} &
\colhead{$\tau_{dyn}$} 
 \\
\colhead{} & 
\colhead{(cm$^{-3}$)} & 
\colhead{(K)} & 
\colhead{(ergs cm$^{-2}$ s$^{-1}$)} & 
\colhead{} & 
\colhead{} &
\colhead{(\kms)} &
\colhead{(\arcsec)} &
\colhead{(pc)} &
\colhead{(kyr)} 
}
\startdata
NGC~2371 & 1260$^a$ & 14,100$^b$ & -10.99$^b$ & 0.11$^c$ & 0.688$^b$ & 42.5$^d$ & 40  & 0.15 & 3.3 \\
Abell~78 & 790$^e$ & 20,800$^b$ & -12.04$^b$ & 0.110$^b$ & 1.61$^b$  & 27$^f$ & 107 & 0.42 & 15\\
NGC~1535 & 2100$^g$ & 10,800$^g$ & -10.45$^b$ & 0.096$^b$ & 0.126$^b$  & 20$^f$ & 33  & 0.13 & 6.3\\
IC~4776 & 12,000$^h$ & 8600$^{b,h}$ & -10.73$^b$ (-10.43$^i$) & 0.085$^b$ & 0.005$^b$  & 22$^j$ & 6   & 0.06 & 2.5\\ 
\enddata
\tablerefs{
(a): \citet{feibelman:97}
(b): \citet{cahn:92}
(c): \citet{kaler:93}
(d): \citet{sabbadin:84}
(e): \citet{medina:00} (inner knots, upper limit)
(f): \citet{weinberger:89}
(g): \citet{tylenda:91}
(h): \citet{phillips:98}
(i): adopted, see text.
(j): \citet{gesicki:00}
}
\end{deluxetable}

\begin{table}[htbp]
\caption{Reddening parameters}\label{tab:red_param}
\begin{tabular}{lccc}
\hline
Star & \EBMV & \EBMV & \EBMV \\
     & (UV slope) & ($c_{H\beta}$) & (\HI) \\
     & (mag)  & (mag) & (mag)\\
\hline
NGC~2371 & 0.070$\pm$0.02 & 0.07$^{a}$ & -\\
Abell~78 & 0.100$\pm$0.02 & 0.12$^{b}$ & 0.21\\
NGC~1535 & 0.055$\pm$0.02 & 0.07$^{b}$ & $0.13\pm0.03$\\
IC~4776  & 0.090$\pm$0.02 & 0.07$^{b}$ & $0.52^{+0.52}_{-0.21}$\\
\hline
\end{tabular}\\
\begin{minipage}{\textwidth}
\scriptsize
\begin{trivlist}
\item a: \citet{kaler:93}
\item b: \citet{cahn:92}
\end{trivlist}
\end{minipage}
\end{table}

\begin{table}[htbp]
\footnotesize
\caption{Derived Stellar Parameters and Adopted Distances}\label{tab:mod_param_dist}
\begin{tabular}{ccccccccccc}
\hline
Star      & \Teff$^{\dagger}$  & \Rtrans$^{\dagger}$ & \vinf$^{\dagger}$ & $D$ & \Rstar$^{\dagger}$ & $\log{L}^{\dagger}$ &
$\log{\Mdot}^{\dagger}$ &$\log{\Mdot}^{\ddagger}$ & $\eta^{\dagger}$ \\
          & (kK)   & (\Rsun) & (\kms) & (kpc) & (\Rsun) & (\Lsun) & (\Msunyr) & (\Msunyr) & \\
\hline
NGC~2371  & $135^{+10}_{-15}$ & $15^{+10}_{-5}$ & $3700\pm200$ & 1.5$^{a}$ &
0.09 &$3.45^{+0.12}_{-0.20}$ & $-7.11\pm0.30$ &  -10.6 & 5.0 \\
Abell~78  & $113\pm8$ & $37^{+20}_{-15}$ & $3200\pm50$ & 1.6$^{a}$ &
0.19 &$3.73^{+0.10}_{-0.13}$ & $-7.33^{+0.36}_{-0.13}$ & -9.3 & 1.4 \\
NGC~1535  & $66\pm5$ & $200^{+100}_{-75}$ & $1950\pm50$ & 1.6$^{b}$&
0.43 & $3.60\pm0.13$ & $-8.10\pm0.30$ & -8.1 & 0.2\\
IC~4776   & $60^{+10}_{-5}$ & $125^{+105}_{-45}$ & $2300\pm200$ & 3.9$^{a}$ &
0.35 & $3.20^{+0.27}_{-0.15}$ & $-7.85\pm0.30$ & -8.8 & 1.0 \\
\hline
\end{tabular}
\begin{minipage}{\textwidth}
\scriptsize
\begin{trivlist}
\item (a): \citet{cahn:92}
\item (b): \citet{sabbadin:84b}
\item $\dagger$: From our spectral modeling.
\item $\ddagger$: Predicted mass-loss rate for our derived parameters
 \Teff, $L$, \logg, \vinf, using prescription of \citet{vink:01}.
\end{trivlist}
\end{minipage}
\end{table}

\begin{table}[htbp]
\footnotesize
\caption{Derived Stellar Parameters From Evolutionary Tracks of \citet{vassiliadis:94}}\label{tab:mod_param_tracks}
\begin{tabular}{cccccc}
\hline

Star & Track & $M_c$ & $M_{init}$ & $\tau_{evol}$ & \logg \\
 &  & (\Msun) & (\Msun) & (kyr) & (\gunit)\\
\hline
NGC~2371  & He & 0.63 & 2 & $\sim13$ & 6.3 \\
Abell~78  & He & $>0.63$ & $>2.0$ & $\lesssim10$ & 5.7 \\
NGC~1535  & H & 0.58 & 1.2 & $\sim15$ & 4.9 \\
IC~4776   & He& 0.57 & 1.0 & $\sim 8$ & 5.1 \\
\hline
\end{tabular}
\end{table}

\begin{table}[htbp]
\caption{Levels and superlevels for model ions}\label{tab:ion_tab}
\footnotesize
\begin{tabular}{lccccccccccc}
\hline
Element & I & II & III & IV & V & VI & VII & VIII & IX & X & XI\\
\hline
H  & 20/30 & 1/1 & & & & & & & & \\
He & 40/45 & 22/30 & 1/1\\
C  & & & 30/54 & 13/18 & 1/1\\
N  & & & & 29/53 & 13/21 & 1/1\\
O  & & & & 29/48 & 41/78 & 13/19 & 1/1\\
Si & & & & 22/33 & 1/1\\
P  & & & & 36/178 & 16/62 & 1/1\\
S  & & & & 51/142 & 31/98 & 28/58 & 1/1\\
Fe & & & & 51/294 & 47/191 & 44/433 & 41/254 & 53/324 & 52/490 &43/210 &1/1\\
\hline
\end{tabular}
\end{table}

\end{document}